\documentclass[aps,prb,showpacs,twocolumn]{revtex4}

\usepackage{graphicx}

\newcommand{\arctanh}{\mathop{\rm arctanh}\nolimits}

\begin{document}
\draft

\title{Magnetization self-organization in a single-domain ferromagnet \\ subject to a spin current}

\author{P.~M.~Gorley$^{1,\, \dag }$, P.~P.~Horley$^1$, V.~K.~Dugaev$^2$, J. Barna\'s$^3$,
 and W.~Dobrowolski$^4$}
\address{
$^1$Department of Electronics and Energy Engineering, Chernivtsi National
University, 2 Kotsyubynsky St., 58012 Chernivtsi, Ukraine\\
$^2$Department of
Physics and CFIF, Instituto Superior T\'ecnico, Av. Rovisco Pais, 1049-001
Lisbon, Portugal, and Frantsevich Institute for Problems of Materials Science,
National Academy of Sciences of Ukraine, 5 Vilde St., 58001
Chernivtsi, Ukraine\\
$^3$Department of Physics, Adam Mickiewicz University,
Umultowska~85, 61-614~Pozna\'n, \\ and Institute of Molecular Physics, Polish
Academy of Sciences, M.~Smoluchowskiego~17, 60-179~Pozna\'n, Poland\\
$^4$Institute of Physics, Polish Academy of Sciences, Al. Lotnik\'ow 32/46,
02668 Warszawa, Poland}
\date{\today }

\begin{abstract}
The Landau-Lifshitz equation for the magnetization dynamics of a
single-domain magnetic system is solved using the methods of
self-organization. The description takes into account the torque
due to spin transfer. The potential energy of the system includes
the uniaxial and easy-plane anisotropies, and the Zeeman energy
due to an external magnetic field. The equilibrium and stationary
states are investigated as a function of the spin current and
external magnetic field. The presented bifurcation diagram allows
to determine the margins of a neutral stability mode of the
equilibrium and stationary states for different values of the
easy-plane anisotropy constant. Using the power spectral density
method, the trajectory tracing, Hausdorff dimension, and maximum
Lyapunov exponent, the dynamics of the phase states in an external
magnetic field is demonstrated. The analytical transcendent
equations for switching between different equilibrium states are
also obtained, proving the importance of phase averaging.
\end{abstract}

\maketitle

\section{Introduction}

Spintronics is a new and intensively developing field of physics,
which attracts significant scientific interest from both
fundamental and application points of view (see, e.g.,
Refs.~[\onlinecite{aRef1,aRef2,aRef3,aRef4,aRef5,aRef6}] and the
references therein). The main objective of spintronics is search
for new materials for devices of future generation (like, for
instance, spin transistors, quantum computing devices, solar
cells, etc.) and with fundamentally new operation principles in
comparison to traditional devices based on the electron charge
transport. \cite{aRef7}$^-$ \cite{aRef9} One of the main problems,
which still require solution, concerns effective control of spin
polarization and spin injection in spintronics devices.
\cite{aRef2, aRef3, aRef10}$^-$ \cite{aRef12} This problem
triggered an extensive theoretical and experimental search for new
materials and devices, like metallic systems consisting of
ferromagnetic and nonmagnetic layers or tunnel junctions including
ferromagnetic and diluted magnetic (non-magnetic) semiconductors.
\cite{aRef1, aRef4}$^-$ \cite {aRef6} Significant progress has
been made recently in the technology and physics of doped
semiconductors with spontaneous spin polarization, and in the spin
injection from magnetic to non-magnetic systems.
\cite{aRef13}$^-$\cite{aInsRef33} Several theoretical models have
been also proposed to describe doping-induced ferromagnetism and
spin injection.\cite{aRef13,aRef14,aRef15,aRef16,aRef17,aRef18,
aRef19,aRef20}

Theoretical studies carried out to date usually treated the spin
subsystem by techniques developed for equilibrium situations.
However, such an approach applied to the spin-transport phenomena
ignores the fact that both spin polarization and spin injection
are dynamical processes and therefore should be treated by methods
of non-equilibrium thermodynamics.\cite{aRef31}$^-$\cite{aRef34}
This is particularly important for open dynamical systems that are
able to exchange particles and energy with the environment (which
is true for the spin subsystem considered here). The dynamical
processes may lead to stationary non-equilibrium states with high
ordering level. Such cooperative phenomena in the system of
magnetic ions and mobile band electrons can take place far from an
equilibrium state. Generally, the system can have several
stationary solutions described by the same set of nonlinear
time-dependent differential equations, which are different
regarding their stability against small fluctuations.

In the case of stable solution, fluctuations are suppressed within a short
period of time. If a solution is unstable, fluctuations can be amplified and
the system may be switched to another stable state. The new state may have the
same or even higher ordering level, when it is described with a lower symmetry
group. Such spontaneous switching between different possible states is known as
a self-organization of the system. The methodology based on non-equilibrium
thermodynamics,  developed for investigating self-organization processes,
offers promising approach to such spintronics problems like spin polarization
and spin injection. This may lead to new quantitative and even qualitative
results, which are not accessible within the traditional approaches.

The problem studied in this paper concerns the dynamics of a
single-domain magnetic system interacting with a spin current
flowing through it -- the problem already investigated in detail
by Sun \cite{aRef20} using the standard methods. The approach based on
the self-organization techniques yields some new interesting
results.

The paper is organized as follows. The model is described in
Section 2, where the symmetry of a quasi-one-dimensional
ferromagnet \cite{aRef20} is analyzed in detail. Analytical
solutions corresponding to four equilibrium and two independent
stationary states are also derived there. The stability analysis
regarding the influence of small perturbations shows that the
spatial angles describing the magnetization vector can vary either
monotonically or in an oscillatory manner with time, depending in
a complex way on the system parameters.

Section 3 presents the results of our numerical calculations and the
relevant discussion. This section is split into two subsections
in order to separate the analysis of equilibrium and stationary
states from the analysis of the time evolution of the system. We also
discuss the magnetization switching and its dependence on the
magnetic field and spin current. It is shown that the equilibrium
and stationary states are characterized by a different dynamics of
the oscillatory modes which can be excited by small perturbations
of the phase variables. The bifurcation points of the system (the
points where the system changes its behavior qualitatively) are
also determined, and their dependence on the external parameters
is investigated. The second subsection presents the results
obtained from the numerical simulations. A significant attention is paid
to the analysis of the system evolution through several types of
different magnetization precession modes. It is shown that the
system can be switched between different equilibrium states,
$m_z=\pm 1$, in a controllable way, which is of particular
interest for the spintronics applications. The detailed analysis of the
system evolution has been performed using the self-organization
methodology; namely, the density histogram of the longitudinal
magnetization component, power spectral density, Hausdorff
dimension, maximum Lyapunov exponent, and the phase trajectory tracing
curve. This allowed us to clarify peculiarities of the
magnetization orientation for different values of the applied
magnetic field. The derived analytical expressions for the magnetic
switching between the states $m_z=\pm 1$ made it possible to
analyze the dependence of the switching time on the external magnetic
field and the spin current. The last Section 4 includes the summary
and main conclusions.

\section{Theoretical model and method}

We consider a model system studied recently by Sun. \cite{aRef20} Accordingly,
we assume a single-domain quasi-one dimensional ferromagnet of length $l_m$
along the axis $x$ and a square cross-section ($a\times a$) in the plane $y-z$.
The system is described by the easy-axis (along the axis $z$) and easy-plane
($y-z$) magnetic anisotropies. In addition, an external magnetic field $\bf H$
is applied in the plane $y-z$ at an angle $\psi$ with respect to the axis $z$.
The external field $\bf H$ and the easy-plane anisotropy are described in
dimensionless relative units, ${\bf h}={\bf H}/H_k$ and $h_p=K_p/K$, where
$H_k=2K/M$, $K$ and $K_p$ are respectively the easy-axis and easy-plane
anisotropy constants, and $M$ is the absolute value of magnetization.

\begin{figure}\label{Fig1}
\includegraphics[scale=1.2]{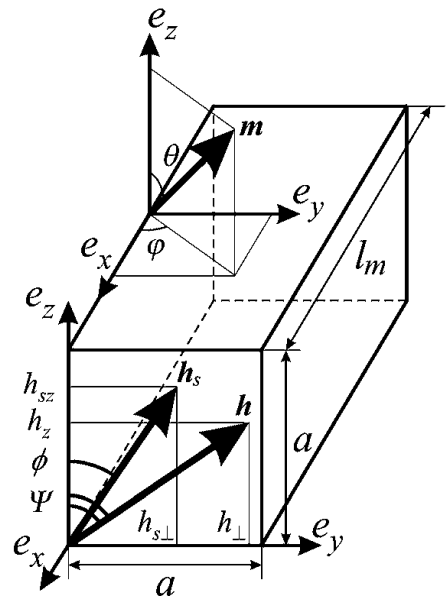}
\caption{ Model geometry and main variables of the system.}
\end{figure}

The system interacts with a spin current ${\bf J}_s$ flowing along
the wire. The incident spin current is described by the two
parameters, $\eta$ and ${\bf n}_s$. The parameter $\eta$ describes
the degree of spin polarization of the incoming charge current
$I$, whereas ${\bf n}_s$ is a unit vector along the corresponding
spin polarization. We assume that ${\bf n}_s$ is in the $y-z$
plane, and forms an angle $\phi$ with the axis $z$. As in
Ref.~[\onlinecite{aRef20}], the spin current is described in
dimensionless units as ${\bf h}_s={\bf n}_s\hbar\eta I/4el_ma^2K$.
In the following we restrict considerations to homogeneous
dynamics, and assume that the system absorbs the perpendicular to
the magnetization component of the incident spin current. This
produces an additional torque acting on the system magnetization.
The geometry of the system, together with the reference frame used
in the theoretical description, are shown in Fig.~1.

As in Ref.~[\onlinecite{aRef20}], we assume that the energy density $U$
includes the energy of uniaxial magnetic anisotropy, $U_K=K\sin ^2\theta$, the
energy of easy-plane anisotropy, $U_p=K_p\, (\sin^2\theta \cos^2 \varphi -1)$, and
the Zeeman energy due to an external magnetic field, ${U_H=-K\, (h_\bot \sin
\theta \sin \varphi +h_z\cos \theta )}$, where $h_\bot =h\sin \psi $ and
$h_z=h\cos \psi $ are the components of the vector $\bf h$ (see Fig.~1). Thus,
one can write
\begin{eqnarray}
\label{Eq1} U_0(\theta,\varphi) \equiv \frac{U(\theta,\varphi)}{K}= Z(\varphi)
\, \sin^2\theta \nonumber \\ -2\,(h_z\cos \theta+h_\bot \sin\theta \sin
\varphi),
\end{eqnarray}
where
\begin{equation}
\label{EqZf} Z(\varphi) \equiv (1+h_p\cos^2\varphi),
\end {equation}
and a constant term in Eq.~(1) has been omitted.

The homogeneous dynamics of the magnetization $\bf{M}$ is
determined by the Landau-Lifshitz equations,\cite{aRef35}
including also the torque due to the spin transfer,\cite{aRef20}
\begin{eqnarray}
\label{Eq2}
\frac{\partial\theta}{\partial\tau}=-\sin\theta(\alpha A_1+A_2), \\
\frac{\partial\varphi}{\partial\tau}=\alpha A_2 - A_1, \nonumber
\end{eqnarray}
with
\begin{eqnarray}
\label {Eq3}
A_1=Z(\varphi)\cos\theta+h_z-\frac{1}{\sin\theta}(h_\bot\cos\theta\sin\varphi +
h_{s\bot}\cos\varphi),\\ A_2=\frac12 h_p \sin 2\varphi
+\frac{1}{\sin\theta}(h_\bot \cos\varphi - h_{s\bot}\cos\theta\sin\varphi).
\nonumber
\end{eqnarray}
Here, $h_\bot=h\sin\psi$ and $h_z=h\cos\psi$  are the applied field components,
$h_{s\bot}=h_s\sin\phi$ and $h_{sz}=h_s\cos\phi$ are the components of spin
current (see Fig.~1), whereas $\tau =t/[(1+\alpha^2)/\gamma H_k]$ denotes the
dimensionless time, which depends on the gyromagnetic ratio $\gamma=g \mu_b
/\hbar$, the anisotropy field $H_k$, and the damping coefficient $\alpha$
($\alpha \ll 1$). When writing down the set of Eqs.~(\ref{Eq2}), we assumed
that the magnitude of the magnetization vector is constant, and $g=2$.

It is rather difficult to obtain analytical solution of
Eqs.~(\ref{Eq2}) and (\ref{Eq3}) in a general case. Therefore, in
this paper we consider a particular case, when the external
magnetic field $\bf h$ and the spin current ${\bf h}_s$ are along the
axis $z$ of the reference system shown in Fig.~1 (similarly to the
situation considered in Ref.~[\onlinecite{aRef20}]). Thus, we put
$h_\bot=0$, $h_z=h$, $h_{s\bot}=0$ and $h_{sz}=h_s$. Note that
from now on, $h$ and $h_s$ denote the $z$-components of the
magnetic field and spin-current. Accordingly,  both $h$ and $h_s$
can take either positive or negative values. Consequently,
Eqs.~(\ref{Eq1}) and (\ref{Eq2}) can be rewritten as
\begin{equation}
\label{Eq4} U_0(\theta,\varphi)=Z(\varphi) \sin^2 \theta - 2 h \cos \theta,
\end{equation}
and
\begin{eqnarray}
\label{Eq5} \frac{\partial \theta}{\partial \tau}=-\sin\theta \left\{ \alpha
[Z(\varphi)\cos\theta + h ] + \frac{h_p}{2} \sin 2 \varphi + h_s \right\}
\nonumber,
\\
\frac{\partial \varphi}{\partial \tau} =\alpha \left[ \frac{h_p}{2} \sin
2\varphi + h_s\right] -[Z(\varphi)\cos\theta + h ].
\end{eqnarray}
Equations~(\ref{Eq5}) present a generalization of Eq.~(11) from
Ref.~[\onlinecite{aRef20}] to arbitrary values of the angle $\theta$ (in
Ref.~[\onlinecite{aRef20}] only the case of $|\theta| \ll 1$ was considered).

The system described by Eqs.~(\ref{Eq5}) is invariant with respect to the
following substitutions
\begin{equation}
\label{Eq6} \varphi \to \pi + \varphi,
\end{equation} and
\begin{eqnarray}
\label{Eq7} h \to \-h,\hskip0.3cm h_s \to -h_s, \nonumber \\ \theta \to \pi -
\theta,\hskip0.5cm \\ \varphi \to \pi - \varphi. \hskip0.5cm \nonumber
\end{eqnarray}
The property (\ref{Eq6}) allows us to reduce the interval of the
variable $\varphi$ to $0 \le \varphi \le \pi$, while the relations
(\ref{Eq7}) can be used to verify analytical results, in
particular to check the solutions of the characteristic equations.

\subsection{Equilibrium solutions}

The equilibrium orientation of the magnetic moment $M$ in the
absence of spin current can be found from the
conditions\cite{aRef36}
\begin{eqnarray}
\label{Eq8} \frac{\partial U_0}{\partial \theta} = 2 \sin \theta_e\,
[Z(\varphi_e)\cos \theta_e+h ]=0, \nonumber \\ \frac{\partial U_0}{\partial
\varphi} = -h_p \sin^2 \theta_e \sin 2 \varphi_e = 0.
\end{eqnarray}
>From now on we label the equilibrium solutions with the index
"$e$", and the stationary ones with the index "0".

Assuming that $h_p > 0\, $,\cite{aRef20} we find the equilibrium solutions of
Eqs.~(\ref{Eq8})
\begin{eqnarray}
\label{Eq9} (1)\hskip1.4cm \sin \theta_e = 0, \hskip1cm \varphi_e - {\rm
arbitrary}, \nonumber\\ (2)\hskip1cm \cos \theta_e = -h,\hskip1.8cm \varphi_e =
\pi/2, \\ (3)\hskip1.9cm \cos \theta_e = -\frac {h} {1+h_p}, \hskip0.5cm
\varphi_e = 0. \nonumber
\end{eqnarray}

\begin{figure}\label{Fig2}
\includegraphics[scale=1]{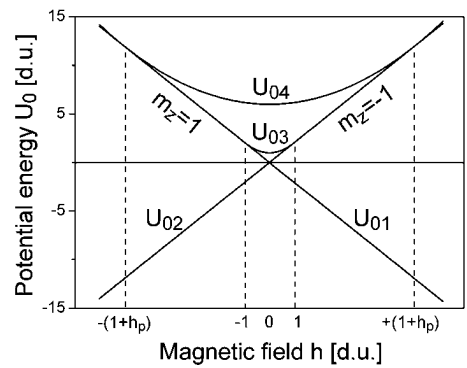}
\caption{ Possible equilibrium states of the system (here and in
the following d.u. means dimensionless units).}
\end{figure}

As follows from Eqs.~(\ref{Eq9}), in the absence of magnetic
field, $h=0$, the system has the following energy states (phases):
spin-degenerate states ${\bf m}=(0,0,\pm 1)$ corresponding to the
energy $U_0=0$, the state ${\bf m}=(1,0,0)$ with the energy
$U_0=1$, and the state ${\bf m}=(0,1,0)$ with the energy
$U_0=1+h_p$. Here, ${\bf m}$ is the unit vector along the
magnetization, ${\bf m}={\bf M}/M$. The external magnetic field
removes the spin degeneracy, and the level $U_0=0$ splits into two
sub-levels with the corresponding energies $U_{01}=-2h$ for the
state ${\bf m}=(0,0,1)$ and $U_{02} = 2h$ for the state ${\bf
m}=(0,0,-1)$. The magnetic field also leads to  precession of the
$m_z$ component around the direction of magnetic field for the
states ${\bf m}=(\sqrt{1-h^2},0,-h)$ with energy $U_{03}=1+h^2$
and ${\bf m}=(0,\sqrt{1-\left({\displaystyle h/(1+h_p)}\right)^2},
-{\displaystyle h/(1 + h_p)})$ with energy
$U_{04}=1+h_p+h^2/(1+h_p)$. The state of energy $U_{03}$ exists in
a certain range of magnetic fields, $0\le |h|\le 1$, while the
state  $U_{04}$ corresponds to $0\le |h|\le 1+h_p$. It is worth
noting that the first solution in Eqs.~(\ref{Eq9}) corresponds to
the minimum of interaction energy for any value of the magnetic
field $h$, while the third solution corresponds to the maximum of
energy for $0 \le |h|\le 1+h_p$ (Fig.~2).

According to Ref.~[\onlinecite{aRef36}], a particular  phase can
be considered stable in a certain range of the field $h$, provided
the following conditions are obeyed:
\begin{eqnarray}
\label{Eq10} \frac{\partial^2 U_0}{\partial \theta^2} > 0,\hskip2cm \nonumber
\\ \frac{2}{\sin^2\theta} \left[ \frac{\partial^2 U_0}{\partial \theta^2}\,
\frac{\partial^2 U_0}{\partial \varphi^2}-\left(\frac{\partial^2 U_0}{\partial
\theta \, \partial \varphi} \right)^2 \right] >0 .
\end{eqnarray}
When one of these expressions turns to zero, with a consequent
change of sign, the corresponding phase becomes unstable.
Substitution of the solutions (\ref{Eq9}) into Eqs.~(\ref{Eq10})
allows one  to determine the conditions of the system stability loss
(for certain values of $\varphi $) for the states, which are odd
with respect to the magnetic field, i.e., for the states $U_{01}$
and $U_{02}$.  The phases $U_{03}$ and $U_{04}$ (even at nonzero
$h$) are unstable in any case. This indicates on the possibility
of spin re-orientation leading to transitions between the existing
phases under applied magnetic fields $0\le h\le 1$ or $1\le |h|\le
1+h_p$.

The boundary corresponding to the transitions between two phases
is defined by the following equations:\cite{aRef36}
\begin{eqnarray}
\label{Eq11} \frac{\partial U_0}{\partial \theta} = 2 \sin \theta \,
[Z(\varphi) \cos \theta + h]=0, \nonumber \\ \frac{\partial^2 U_0}{\partial
\theta^2}=2\, [Z(\varphi) \cos 2 \theta + h \cos \theta]=0.
\end{eqnarray}
As the single solution of Eq.(\ref{Eq11}) coincides with the first
solution in (\ref{Eq9}), it means that $\theta = 0$, $\varphi
=0.5\, \arccos \, [-2(h+1)/h_p-1]$ for $-(1+h_p)\le h\le -1$ with
$U_{01}=-2h$ and $\theta =\pi$, $\varphi =0.5\, \arccos \, [2\,
(h-1)/h_p-1]$ for $1\le h\le 1+h_p$ with $U_{02}=2h$ describe the
lines of the second order phase transition,\cite{aRef36} in
accordance with the Landau theory.\cite{aRef37}

\subsection{Constant energy solutions}

Using Eqs.~(\ref{Eq4}) and (\ref{Eq5}), we find that the variation
of $U_0(\theta,\varphi)$ with time obeys the following equation:
\begin{eqnarray}
\label{Eq12} \frac{\partial U_0(\theta, \varphi)}{\partial \tau} =-2 \sin^2
\theta \left\{ \alpha \left[ (Z(\varphi) \cos \theta + h)^2 \right. \right.
\nonumber \\ \left. \left. +(0.5 h_p \sin 2\varphi +h_s)^2\right] +h_s\left[
Z(\varphi) \cos \theta \right. \right. \nonumber \\ \left. \left. +h-\alpha \,
(0.5 h_p\sin 2\varphi+h_s) \right] \right\}.
\end{eqnarray}
>From this follows that the potential energy of the system is an
integral of motion [$U_0(\theta,\varphi)\equiv U_0={\rm const}$]
when both $\alpha = 0$ and $h_s = 0$, which allows to transform
Eq.~(\ref{Eq5}) to the form
\begin{eqnarray}
\label{Eq13}
\frac{\partial\varphi}{\partial\tau}=\mp \left[
h^2-U_0Z(\varphi)+Z(\varphi)^2\right] ^{1/2}, \nonumber \\ Z(\varphi)\,
\cos\theta +h = -\frac{\partial \varphi}{\partial \tau}\, .\hskip0.5cm
\end{eqnarray}
The "$+$" and "$-$" signs on the right-hand-side correspond to the two possible
solutions for $\cos \theta$ from Eq.~(\ref{Eq4}) at $U_0=$const.

Equations (\ref{Eq13}) have real solutions only when
\begin{eqnarray}
\label{Eq14}
h^2+Z(\varphi)^2 \ge U_0 Z(\varphi),\hskip1cm \nonumber \\ -1 \le
\cos \theta = - \frac1{Z(\varphi)} \left({\displaystyle \frac{\partial
\varphi}{\partial \tau }} +h\right) \le 1.
\end{eqnarray}
The analysis of condition (\ref{Eq14}) shows that the solutions of
Eqs.~(\ref{Eq13}) exist for $|2h| < U_0 < U_{03}$ and $U_{03} <
U_0 < U_{04}$, when it can be expressed by the Jacoby elliptic
functions.\cite{aInsRef34} However, if the interaction energy
$U_0$ coincides with the energy of one of the equilibrium states
(\ref{Eq9}), the solutions can be simplified to the inverse
trigonometric or exponential functions. In particular, for the
case of $U_0=U_{01,2}=\pm 2 |h|$, the solutions of
Eqs.~(\ref{Eq13}) are
\begin{eqnarray}
\label{Eq15}
\varphi = \mp \arctan \left[ \sqrt{\xi / \zeta} \; \tanh
(\sqrt{\zeta\xi}\, \tau) \right] ,\hskip1cm \nonumber \\ \cos \theta = \pm
\left[ 1 - 2|h| \frac{\zeta+\xi \tanh^2
(\sqrt{\zeta\xi}\tau)}{\zeta(1+h_p)+\xi\tanh^2(\sqrt{\zeta\xi}\tau)}\right],\\
\xi \equiv (1+h_p - |h|), \hskip0.5cm \zeta \equiv |h|-1. \nonumber
\end{eqnarray}
Here, the upper sign corresponds to the interval of $1<h\le 1+h_p$, while the
lower one is for the interval of $-(1+h_p)\le h<-1$. When deriving
Eq.~(\ref{Eq15}) we assumed  the integration constant $\tau_0=0$. Note that for
$|h|\to 1$, the values of $\varphi$ and $\cos\theta$ tend to the finite limits
\begin{eqnarray}
\label{Eq16} \varphi_{|h| \to 1}=\mp \arctan \, (h_p \tau),\hskip1cm \nonumber
\\ \hskip1cm \cos \theta_{|h| \to 1} =\pm \left( -1 + \frac {2 h_p}{1+h_p
+h_p^2 \tau^2}\right) .
\end{eqnarray}

It is important to note, that apart from the states described by
Eq.~(\ref{Eq15}), there also exists the state with $\cos\theta =
-1$ for arbitrary $\varphi$ and $h > 0$, and the state with
$\cos\theta = 1$ for $h<0$. As follows from (\ref{Eq16}), if we
neglect the plane anisotropy ($h_p=0$) or take the limit of $\tau
\to \infty$, Eq.~(\ref{Eq15}) for $\cos\theta$ gives the
independent on $\varphi$ expressions mentioned above. In other
words, Eqs.~(\ref{Eq15}) and (\ref{Eq16}) prove that along the
lines corresponding to the phase transitions of the second type,
there are two possible solutions for the longitudinal
magnetization component: a constant one, and a periodic solution
with a complex time dependence. In particular, for $|h|>1$ it is
characterized by the cyclic frequency
\begin{equation}
\label{Eq17}
\omega(h_p,h)= \left[ \left( |h|-1\right) \left( 1+h_p-|h|\right)
\right] ^{1/2}.
\end{equation}
It is worth noting that Eq.~(\ref{Eq15}) for $\varphi(\tau)$ and
Eq.~(\ref{Eq17}) almost coincide with similar expressions for $\varphi(\tau)$
and frequency $\omega_p$ from Ref.~[\onlinecite{aRef20}], if one transforms the
field interval from $h>-1$ to $1 \le |h| \le 1+h_p$ considered here.

When $U_0=U_{03,4}$ the solution of (\ref{Eq13}) has the form:

a) for the case $U_0=U_{03}$:
\begin{eqnarray}
\label{InsEq19} \sin^2\varphi = \left[ k^{\prime 2} \cot^2 \left(k^\prime h_p
(\tau - \tau_0)/k \right) - k^2 \right]^{-1},\\ Z(\varphi)\cos \theta_{1,2}=-h
\pm \sqrt{(Z(\varphi)-1)(Z(\varphi)-h^2)}, \nonumber
\end{eqnarray}

where $k^2=h_p/(1+h_p-h^2)$ and $k^{\prime 2} = 1-k^2$.

b) for the case $U_0=U_{04}$ and $h>\sqrt{1+h_p}$:
\begin{eqnarray}
\label{InsEq20} \cos^2\varphi = (A^2-1) \frac {\tan^2 \left[
h_p(\tau-\tau_0)/A \right]}{A^2 \tan^2 \left[
h_p(\tau-\tau_0)/A \right] +1 }, \nonumber \\
Z(\varphi)\cos \theta_{1,2}=-h \pm (1+h_p)^{-1/2}
\\ \times \sqrt{(1+h_p-Z(\varphi)) (h^2- (1+h_p) Z(\varphi))}, \nonumber
\end{eqnarray}
where $A^2 = h_p (1+h_p) / \left[ (1+h_p)^2 - h^2 \right] >1$, and the
expression for $Z(\varphi)$ is given by Eq.~(\ref{EqZf}). The integration
constant $\tau_0$ in Eqs.~(\ref{InsEq19}) and (\ref{InsEq20}) can be defined
from the initial conditions. The comparison of formulas (\ref{Eq16}),
(\ref{InsEq19}) and (\ref{InsEq20}) proves that the magnetization dynamics
corresponding to the equilibrium energy is expressed with the complex functions of
$h$ and $h_p$, depending on the form of solutions (\ref{Eq9}).

\subsection{Analytical solutions in the presence of spin current}

Let us now turn back to the discussion of the effect of spin current. It is
quite natural to expect that the spin current can affect the system in a way
similar to that of an external magnetic field. In a general case, the influence
of spin current on the magnetization dynamics can be studied only by the numerical
integration of Eqs.~(\ref{Eq5}). This will be presented later. Below we
consider three particular cases of $h_s \ne 0$, for which some analytical
solutions can be derived.

The simplest case is the one with $\alpha = 0$ and $h=0$, when Eqs.~(\ref{Eq5})
transform to
\begin{eqnarray}
\label{Eq18}
\frac{\partial \theta}{\partial \tau} =-\left( \frac12 \, \sin 2
\varphi + h_s\right) \sin \theta, \nonumber \\ \frac{\partial \varphi}{\partial
\tau}= -Z(\varphi) \, \cos \theta .\hskip1cm
\end{eqnarray}

Excluding the time variable, the above system of two equations can be reduced
to a single equation with the corresponding solution
\begin{equation}
\label{Eq19} \theta =\arcsin \left[ U_0(\varphi) / Z(\varphi)\right] ^{1/2},
\end{equation}
where
\begin{equation}
\label{Eq21} U_0(\varphi)=\exp \left( -\frac{2 h_s \arctan(\sqrt{1+h_p}\cot
\varphi) }{\sqrt{1+h_p}}\right).
\end{equation}
The integration constant in Eq.~(\ref{Eq19}) was determined from the condition
that $\theta =\pi /2$ corresponds to $\varphi =\pi/2$.

Substituting Eq.~(\ref{Eq19}) into Eqs.~(\ref{Eq18}) and (\ref{Eq4}), we obtain
\begin{equation}
\label{Eq20} \frac{\partial\varphi}{\partial\tau} =-Z(\varphi) \left[
1-\frac{U_0(\varphi)}{Z(\varphi)}\right] ^{1/2}.
\end{equation}
It is obvious that the solution of Eq.~(\ref{Eq20}) can be found
only by numerical methods. However, some analytical results are
achievable regarding the magnitude and sign of $h_s$. Equation
(\ref{Eq20}) can have real solutions only when
\begin{equation}
\label{Eq22} \frac {\sqrt{1+h_p} \; \ln Z(\varphi)} {2 \arctan \,
(\sqrt{1+h_p}\, \cot \varphi )} \ge -h_s.
\end{equation}

If $h_s>0$, the condition (\ref{Eq22}) is automatically fulfilled. With
increasing spin current, the variables $\varphi(\tau)$, $\theta(\tau)$ and
$U_0(\tau)$ relax to their equilibrium values, corresponding to the limiting
case of $h_s \to \infty$~:
\begin{eqnarray}
\label{Eq23} \varphi = -\arctan \left[ \sqrt{1+h_p}\, \cot \left(
\sqrt{1+h_p}\, \tau-\frac{\pi}{2}\right) \right], \\ \sin \theta =0,
\hskip0.5cm U_0=0. \nonumber
\end{eqnarray}
The integration constant for $\varphi(\tau)$ in Eq.~(\ref{Eq23}) was chosen to
keep $\varphi (\tau =0)=0$. It is important to note that the nature of the
solution (\ref{Eq23}) coincides with the first solution in Eq.~(\ref{Eq9}).

For $h_s<0$ and for certain values of spin current,
Eq.~(\ref{Eq22}) is no longer valid and the system cannot be
integrated. When the left and right sides of Eq.~(\ref{Eq22}) are
equal, the system exists in the state
\begin{equation}
\label{Eq24} \sin\theta = 1,\hskip0.5cm \varphi = 0,\hskip0.5cm U_0=1,
\end{equation}
which coincides with the second solution in Eq.~(\ref{Eq9}).

Therefore, when no magnetic field is applied, the spin current can
excite the magnetic system, which in turn may relax either to the
first or to the second state described by Eqs.~(\ref{Eq9}),
depending on the sign of $h_s$. In other words, the behavior of the
system depends on both direction and magnitude of the spin
current. One has to pay special attention to the exponential
character of this dependence, because a comparatively small change
of $h_s$ in certain direction can result in a significant change
of the magnetization orientation and of the interaction energy. It
is also worth noting that the dependence of the magnetization
components and the interaction energy on the applied magnetic field is
much weaker, being linear or quadratic in $h$.

The second case, where some analytical solutions are possible, is the situation
with  $h_p=0$. Equations (\ref{Eq5}) can be then rewritten for $m_z=\cos\theta$
as
\begin{eqnarray}\label{Eq25}
\frac{\partial m_z}{\partial \tau}=\alpha (1-m_z^2)(m_z+h+h_s/\alpha) \\
\frac{\partial \varphi}{\partial \tau}=-(m_z+h-\alpha h_s) \nonumber
\end{eqnarray}
It can be shown that for the case $\alpha h + h_s \ne \pm \alpha$, the solution
of Eq.~(\ref{Eq25}) has the following form
\begin{eqnarray}
\label{Eq26}
\frac{1-m_z^2}{(m_z+\beta)^2}\;\left(\frac{1+m_z}{1-m_z}\right)^\beta
\nonumber \\
=\frac{1-m_{z0}^2}{(m_{z0}+\beta)^2} \; \left(
\frac{1+m_{z0}}{1-m_{z0}} \right)^\beta \; e^{-2 \alpha (1-\beta^2)\tau}, \\
\varphi(\tau)= \varphi_0 - (h-\alpha h_s)\,
\tau - \int \limits_0^\tau m_z(\tau)\, d\tau, \nonumber
\end{eqnarray}
where $\varphi(0)$ is the initial value of the angular variable and
\begin{equation}
\label{Eq27} \beta = h + \frac{h_s}{\alpha}.
\end{equation}

Equations (\ref{Eq26}) and (\ref{Eq27}) represent a generalization of Eqs.~(22)
and (23) from Ref.~[\onlinecite{aRef20}] to the case of arbitrary $\theta$. As
follows from Eq.~(\ref{Eq26}), in the stationary case ($\tau \to \infty$) and
for vanishing easy-plane anisotropy, the system turns to the state with $m_z =
\cos \theta_0 = -1$ in a complicated manner. At the initial time, the
interaction energy has a jump $\left. \partial U_0(h_p=0) /
\partial \tau \right|_{\tau=0}=-2 \alpha \beta h$; further on, depending on
the sign of $\beta$, it increases or decreases in an oscillatory manner to
reach the value of $\left. \partial U_0 (h_p=0) / \partial \tau \right|_{\tau
\to \infty}=0$.

Some analytical solutions can also be obtained for the stationary situations, when
Eqs.~(\ref{Eq5}) can be transformed into a trigonometric form with the general
periodic solutions $\theta_0$ and $\varphi_0$ given by
\begin{eqnarray}
\label{Eq28} \sin 2\varphi_0 = - 2 h_s / h_p, \nonumber \\ \cos\theta_0 = - h /
Z(\varphi_0).
\end{eqnarray}
The invariance conditions (\ref{Eq6}) allow to select only two independent
stationary states from those described by Eq.~(\ref{Eq28}):
\begin{eqnarray}
\label {Eq29} \varphi_{01}+\varphi_{02} =\pi/2, \hskip1.5cm \nonumber
\\
\cos \varphi_{02} =-{\rm sign}\,  h_s \sqrt{\frac12 \left( 1-\sqrt{1-4 h_s^2 /
h_p^2}\right)},
\\
\sin \varphi_{02} =\sqrt{\frac12 \left(1+\sqrt{1- 4 h_s^2 /h_p^2}\right)},
\nonumber \\ \sin \varphi_{01} =\cos \varphi_{02}, \hskip0.5cm
\cos\varphi_{01}=\sin\varphi_{02}, \nonumber \\
\cos\theta_{01,2}=-h/Z(\varphi_{01,2}). \nonumber
\end{eqnarray}

As follows from Eqs.~(\ref{Eq29}), the stationary states are possible only for
$h_s$ satisfying the condition $|h_s| \le 0.5 h_p$. At the same time, the
positive or negative value of the longitudinal magnetization vector component
is determined by the orientation of external magnetic field, being independent
of the angle between the spin current and $e_z$. It is important to note that
the stationary states described by Eq.~(\ref{Eq29}) are characterized by the
potential energy of Eq.~(\ref{Eq4}), which is an integral of motion with the
corresponding eigenvalues
\begin{equation}
\label{Eq30}
U_{S01,2}=Z(\varphi_{01,2})+h^2/Z(\varphi_{01,2}).
\end{equation}
The differences between the equilibrium energies $U_{03}=1+h$ and
$U_{04}=1+h_p+h^2/(1+h_p)$ and these stationary states $U_{S01,2}$ are
\begin{eqnarray}
\label{Eq31}
\left.\Delta U_{01}\right|_{0 \le |h| \le 1+h_p}=U_{04}-U_{S01}
\hskip2cm \nonumber \\ =y - \frac{y h^2}{(1+h_p)(1+y + h_p \sqrt{1-x^2})} ,
\nonumber \\ \left. \Delta U_{02} \right|_{0\le |h| \le 1} = U_{S02}-U_{03} =y
- \frac{y h^2}{1+y} ,\hskip0.5cm \\ x \equiv \left| \frac{2 h_s}{h_p} \right|,
\hskip0.5cm y\equiv \frac{1}{2}~h_p \left( 1 - \sqrt{1-x^2} \right).\nonumber
\end{eqnarray}
Therefore, Eqs.~(\ref{Eq29}) to (\ref{Eq31}) demonstrate that neither
stationary states nor the corresponding energies depend on the damping
coefficient $\alpha$, as they are only determined by $h$, $h_p$ and $h_s$. From
Eq.~(\ref{Eq31}) follows that for $h_s=0$, the stationary state $U_{S01}$
coincides with $U_{04}$, and $U_{S02}$ with $U_{03}$. Under the influence of
spin current, the state $U_{04}$ shifts towards the lower energy and becomes
more thermodynamically stable transforming into the state $U_{S01}$, while the
state $U_{03}$ increases in energy and transforms into the state $U_{S02}$. As
follows from Eq.~(31), the relative position of both stationary states on the
energy scale could be then changed in controllable way by means of a proper
choice of $h$ and $h_s$.

\subsection{Stability of the equilibrium and stationary states}

To investigate the stability of equilibrium (\ref{Eq9}) or stationary
(\ref{Eq29}) solutions, we subject them to small perturbations
\begin{equation}
\label{Eq32} \theta=\theta_i+\delta\theta_i, \quad \varphi=\varphi_i
+\delta\varphi_i , \hskip0.5cm (i=e {\rm \; or \; } 0)
\end{equation}
with $\delta\theta_i \ll \theta_i$, $\delta\varphi_i \ll \varphi_i$.
Physically, such perturbations can have quite different origin: they can be
caused either by a non-homogeneity of the sample, presence of contacts,
fluctuations of the temperature or external fields,\cite{aRef38} etc.
Substituting (\ref{Eq32}) into (\ref{Eq5}) and performing a standard
linearization procedure over small perturbations,\cite{aRef31} one obtains the
dynamic matrix \textbf{\emph{a}} with the corresponding eigenvalues
\begin{equation}
\label{Eq33} \lambda_{1,2}=0.5\left(
a_{11}+a_{22}\pm\sqrt{(a_{11}-a_{22})^2+4a_{12}a_{21}}\right)
\end{equation}
and the matrix elements $a_{ij}$
\begin{eqnarray}
\label{Eq34} a_{11}=-\cos\theta_1 \{ \alpha [ Z(\varphi_i) + h] \nonumber + 0.5
h_p \sin 2 \varphi_i + h_s\}
\\
+ \alpha Z(\varphi_i) \sin^2 \theta_i, \nonumber \\ a_{12} =h_p\sin\theta_i
(\alpha\cos\theta_i\sin2\varphi_i-\cos2\varphi_i),\hskip0.5cm \\
a_{21}=\sin\theta_iZ(\varphi_i),\hskip2cm \nonumber \\ a_{22} =
h_p(\cos\theta_i\sin2\varphi_i+\alpha\cos2\varphi_i).\hskip1cm \nonumber
\end{eqnarray}

After substituting the solutions (\ref{Eq9}) into Eq.~(\ref{Eq34}) and then
into Eq.~(\ref{Eq33}), one obtains three different solutions:

a) for the state with $\sin\theta_e = 0$, $\varphi_e$ arbitrary:
\begin{eqnarray}
\label{Eq35} \lambda_1=\mp \{ \alpha [Z(\varphi_e)+h] + 0.5 h_p \sin 2
\varphi_e + h_s\},
\\
\lambda_2 = h_p (\pm \sin 2 \varphi_e + \alpha \cos 2 \varphi_e),\hskip1cm
\nonumber
\end{eqnarray}
with the upper and lower signs corresponding to $\theta_e = 0$ and $\theta_e
=\pi$, respectively.

b) for the state with $\cos \theta_e = -h$, $\varphi_e = \pi/2$
\begin{eqnarray}
\label{Eq36} \lambda_{1,2}=\frac{1}{2}\left[ h \, h_s + \alpha (1+h-h_p)
\right. \hskip1cm
\\
\left. \pm \sqrt{[h \, h_s + \alpha (1+h+h_p)]^2+4 h_p (1-h^2)} \right].
\nonumber
\end{eqnarray}

c) for the state with $\cos \theta_e = -h/(1+h_p)$, $\varphi_e = 0$:
\begin{eqnarray}
\label{Eq37} \lambda_{1,2}=\left.\frac{0.5}{1+h_p}\right[ h h_s + \alpha
(1+h_p) (1+2 h_p + h) \nonumber \\ \pm \{[ h h_s +\alpha (1+h_p)(1+h)]^2 \\ -
\left. 4 h_p (1+h_p) [(1+h_p)^2 - h^2] \}^{1/2} \right], \nonumber
\end{eqnarray}

The solutions of the characteristic equation, corresponding to the
stationary states (\ref{Eq29}), are rather complicated and we have
calculated them numerically. It is worth noting that the solutions
(\ref{Eq35}) and (\ref{Eq36}) are real ($\lambda_1>0$ and
$\lambda_2<0$) -- also for the stationary state ($\theta_{02}$,
$\varphi_{02}$). In turn, the solutions (\ref{Eq37}) are
complex-conjugated ($\lambda_{1, 2} = \lambda_r \pm
i\lambda_{im}$), including the case of stationary state
($\theta_{01}$, $\varphi_{01}$).

The stability of the equilibrium or stationary states can be estimated using
different stability criteria. In our case, we apply the method of Lyapunov,
\cite{aRef39} evaluating the stability from the signs of $\lambda_1$ and
$\lambda_2$, as well as the criteria of Gurvitz,\cite{aRef40, aRef41} taking
into account the relations between the matrix elements $a_{ij}$. In particular,
according to Gurvitz, the sufficient conditions for stability of the system
under consideration are
\begin{equation}
\label{Eq38} a_{11}+a_{22}<0, \qquad a_{11}a_{22}>a_{12}a_{21}.
\end{equation}
If $\lambda_1$ and $\lambda_2$ (see Eq.~(\ref{Eq33})) are known, the solutions
$\theta(\tau)$ and $\varphi(\tau)$ have the following form \cite{aRef39}

(a) for the case of real $\lambda_1$ and $\lambda_2$:
\begin{eqnarray}
\label{Eq39} \theta(\tau)=\theta_i+C_{11}e^{\lambda_1\tau
}+C_{12}e^{\lambda_2\tau } , \nonumber \\
\varphi(\tau)=\varphi_i+C_{21}e^{\lambda_1\tau }+C_{22}e^{\lambda_2\tau },
\end{eqnarray}

(b) for the complex-conjugate $\lambda_1$ and $\lambda_2$:
\begin{eqnarray}
\label{Eq40} \theta(\tau) =\theta_i +2\left[
D_{11}\cos(\lambda_{im}\tau)-D_{12}\sin(\lambda_{im}\tau)\right]
e^{\lambda_r\tau},\hskip0.3cm \nonumber
\\
\varphi(\tau)=\varphi_i +2\left[ D_{21}\cos (\lambda_{im}\tau)-D_{22}\sin
(\lambda_{im}\tau)\right] e^{\lambda_r\tau}.\hskip0.3cm
\end{eqnarray}
Here $C_{ij}$ and $D_{ij}$ are the integration constants, which are generally
determined from the relevant boundary conditions. The latter ones can be
written as
\begin{eqnarray}
\label{Eq41} \left.\theta(\tau)\right|_{\tau=0} = \theta_i + 2 \pi, \hskip0.5cm
\left.\varphi(\tau)\right|_{\tau = 0}= \varphi_i + \pi, \nonumber
\\ \left.\frac{\partial \theta(\tau)}{\partial \tau}\right|_{\tau=0} = 2 \left.
\frac {\partial \varphi(\tau)}{\partial \tau}\right|_{\tau=0} = -2 \pi
|\lambda_2|
\end{eqnarray}
for real $\lambda_1$ and $\lambda_2$, and
\begin{eqnarray}
\label{Eq42} \left.\theta(\tau)\right|_{\tau=0} = \theta_i, \hskip1cm
\left.\varphi(\tau)\right|_{\tau = 0}= \varphi_i, \\ \left.\frac{\partial
\theta(\tau)}{\partial \tau}\right|_{\tau=0} = 2 \left. \frac {\partial
\varphi(\tau)}{\partial \tau}\right|_{\tau=0} = -2 \pi |\lambda_{im}| \nonumber
\end{eqnarray}
for complex-conjugate $\lambda_1$ and $\lambda_2$. Equations~(\ref{Eq39}) and
(\ref{Eq40}) can be rewritten as
\begin{eqnarray}
\label{Eq43} \theta_2(\tau)=\theta_{i2}+2\pi \exp (\lambda_2 \tau), \\
\varphi_2(\tau)=\varphi_{i2}+\pi \exp (\lambda_2 \tau) \nonumber
\end{eqnarray}
for real $\lambda_1$ and $\lambda_2$, and
\begin{eqnarray}
\label{Eq44} \theta_1(\tau)=\theta_{i1}+2\pi \exp (\lambda_r \tau) \, \sin
(\lambda_{im} \tau), \\ \varphi_1(\tau)=\varphi_{i1}+\pi \exp (\lambda_r
\tau)\, \sin (\lambda_{im} \tau) \nonumber
\end{eqnarray}
for complex conjugated $\lambda_1$ and $\lambda_2$. As follows
from Eqs.~(\ref{Eq43}) and (\ref{Eq44}), the functions
$\theta(\tau)$ and $\varphi(\tau)$ can have either monotonic or
oscillatory character, with the characteristics depending on the
control parameters in a rather complex way, which can be
investigated by numerical calculations described later.

\subsection{Phase averaging}

Let us come back to Eq.~(\ref{Eq12}) describing the time variation
of the interaction energy. Using Eqs.~(\ref{Eq5}), one can rewrite
this equation in the form
\begin{eqnarray}
\label{Eq45} 2 h_s \sin^2 \theta \; \frac {\partial \varphi}{\partial \tau} -
\frac{\partial U_0}{\partial \tau} = 2 \alpha \sin^2 \theta \hskip1.5cm
\nonumber \\ \times \{ [Z(\varphi) \cos \theta + h]^2 + (0.5h_p \sin 2 \varphi
+h_s)^2\}.
\end{eqnarray}
On the right side of Eq.~(\ref{Eq45}) there is a positive
quadratic form with a constant magnitude under any transformation
of the coordinate system. Hence, when the spin current is absent,
the energy $U_0(\theta,\varphi,\tau)$ decreases with time, i.e.,
the system tends to an equilibrium state. If $h_s \ne 0$, the
variation of energy $\partial U_0 / \partial \tau$ can be either
positive or negative, depending on the sign and magnitude of the
first term on the left side of Eq.~(\ref{Eq45}). As $\partial
\varphi / \partial \tau$ is determined by $h_s$ for the fixed
values of $h_p$ and $\alpha$, it is natural that $\partial U_0 /
\partial \tau$ also depends mainly on the direction and magnitude
of the spin current.

To reveal the exact form of this dependence, one has to average
Eq.~(\ref{Eq45}) over the phase. When performing such an averaging,
we take into account the periodicity of Eq.~(\ref{Eq45}) and also the
invariance relations (\ref{Eq6}). For the averaging procedure we
will use two possible approaches: without any weighting
coefficients\cite{aRef40}
\begin{equation}
\label{InsEq47} \langle f \rangle \equiv \langle f(\theta, \varphi) \rangle =
\frac{1}{\pi^2} \int\limits_0^\pi \int\limits_0^\pi f(\theta, \varphi) d\theta
d\varphi,
\end{equation}
and with weighting coefficients\cite{aRef40}$^,$ \cite{aInsRef35}:
\begin{equation}
\label{InsEq48} \langle f \rangle \equiv \langle f(\theta, \varphi) \rangle =
\frac{1}{2\pi} \int \limits_0^\pi \int \limits_0^\pi f (\theta, \varphi) \sin
\theta d\theta d\varphi.
\end{equation}

Substituting (\ref{Eq45}) into (\ref{InsEq47}) and (\ref{InsEq48}), one
obtains

a) for the averaging procedure (\ref{InsEq47})
\begin{equation}
\label{InsEq49} -\left[ h \frac{h_s}{\alpha} + \frac{1}{\alpha} \left\langle
\frac {\partial U_0}{\partial \tau} \right\rangle \right] = h^2 +
\frac{1}{4}\left( 1+\frac{h_p}{2}\right) ^2+\frac{5h_p^2}{32},
\end{equation}

b) for the averaging procedure (\ref{InsEq48})
\begin{equation}
\label{InsEq50} - \left[ \frac{4 h\cdot h_s}{3 \alpha} + \frac{1}{\alpha}
\left\langle \frac{\partial U_0}{\partial \tau} \right\rangle \right] =
\frac{4}{3}\left[ h^2 + \frac{(1+h_p+h_p^2)}{5}\right].
\end{equation}

First of all, it is necessary to emphasize that the expressions
(\ref{InsEq49}) and (\ref{InsEq50}) have the same functional
dependence on $h$ and $h_s$, correlating well with the invariance
upon simultaneous replacement $h \to -h$ and $h_s \to -h_s$ (see
Eq.~(\ref{Eq7})). As one can see, the right hand sides of the
obtained formulas do not include $h_s$. This means that when the
external magnetic field and the spin current have the same direction,
the average energy can decrease faster than in the case of
$h_s=0$. For antiparallel $h$ and $h_s$, the average  $\langle
\partial U_0 /
\partial \tau \rangle$ can be negative, zero, or can become
positive -- but its absolute value is always smaller than $|h
\cdot h_s|$ to obey Eqs. (\ref{InsEq49}) or (\ref{InsEq50}). It is
worth noting that the approximation $\theta \ll 1$, used in
[\onlinecite{aRef20}], leads to the violation of the symmetry
condition (\ref{Eq7}), which is also reflected in the averaging
formula (\ref{InsEq47}) for  $\langle \partial U_0 /
\partial \tau \rangle$:
\begin{eqnarray}
\label{InsEq51} - \left[ \frac{h_s}{\alpha} \left(1+\frac{h_p}{2}+h
\right)+\frac{1}{2 \alpha \theta_0^2} \left\langle \frac{\partial U_0}{\partial
\tau} \right\rangle_{|\theta \ll 1} \right] \\ = \left( 1+ \frac{h_p}{2} + h
\right)^2 + \frac{h_p^2}{4}. \nonumber
\end{eqnarray}
The averaging procedure (\ref{InsEq48}) in the given approximation
should be changed to
\begin{equation}
\label{InsEq52} \langle f \rangle_{|\theta \ll 1} = \frac{1}{\pi \theta_0} \int
\limits_0^{\theta_0} \int \limits_0^\pi \theta f(\theta, \varphi) d\theta
d\varphi,
\end{equation}
leading to two times greater coefficient at $\langle \partial U_0 /
\partial \tau \rangle$ in (\ref{InsEq51}).

The difference between the expression (\ref{InsEq51}) and the
corresponding formula (\ref{Eq15}) from Ref. [\onlinecite{aRef20}]
is caused by the fact that we take into account the quadratic term
in the expansion $\left. \cos \theta \right|_{\theta \ll 1}
\approx 1 - \theta^2/2$, which was neglected in
Ref.~[\onlinecite{aRef20}].

\section{Results of numerical calculations and their discussion}

Using the formulas presented above, the  magnetization dynamics of
a single-domain ferromagnet can be investigated as a function of
the control parameters: $\alpha$, $h_p$, $h_s$ and $h$. In our
calculations, the number of control parameters was reduced to $h$
and $h_s$ by assuming $\alpha=0.005$ (see
Ref.~[\onlinecite{aRef11}]) and $h_p=5$ (as in
Ref.~[\onlinecite{aRef20}]).

\subsection{Analysis of equilibrium and stationary states}

Let us discuss first the dependence of stationary magnetization
states on the parameters $h$ and $h_s$. Figure 3 shows the
magnetic field dependence of $m_x=\sin\theta_0\cos\varphi_0$,
$m_y=\sin\theta_0\sin\varphi_0$, and $m_z=\cos\theta_0$ components
of a unit magnetization vector ${\bf m}$, corresponding to the
stationary states (\ref{Eq29}). All other control parameters are
kept constant in the calculations, namely $\alpha=0.005$, $h_p=5$
and $h_s=-0.03$. As one can see, there is a range of magnetic
field, where the magnetization components are non-zero for both
stationary states. It is important to emphasize that the limiting
field values $\pm h_{lim}$ depend on $h_s / h_p$ for the
stationary state ($\theta_{01}$, $\varphi_{01}$), remaining
practically independent on $h_s / h_p$ for the state
($\theta_{02}$, $\varphi_{02}$) with $h_{lim}=\pm1$. In both
cases, the longitudinal component $m_z$ varies linearly with
magnetic field, whereas the transverse components $m_x < m_y$ have
a non-linear field dependence. For better visual presentation of the
data, the curves 3 and 5 in Fig.~3 have been multiplied by the
factor of 40.

\begin{figure}\label{Fig3}
\hspace*{-0.5cm}
\includegraphics[scale=1]{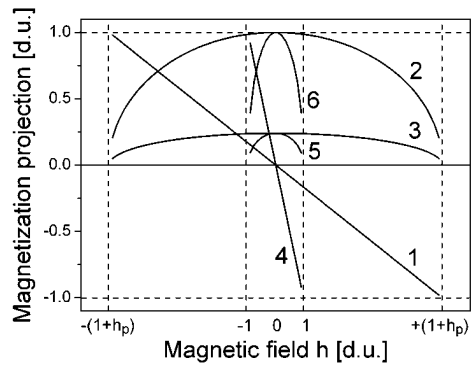}
\caption{ Field dependence of the stationary magnetic moment
components. The curves labelled with 1, 2 and 3 correspond
respectively to $m_{z1}$, $m_{x1}$ and $40\, m_{y1}$; whereas the
curves 4, 5 and 6 represent $m_{z2}$, $40\, m_{x2}$, and $m_{y2}$.
(The curves 3 and 5 have been multiplied by the factor of 40.) }
\end{figure}

\begin{figure}\label{Fig4}
\includegraphics[scale=0.8]{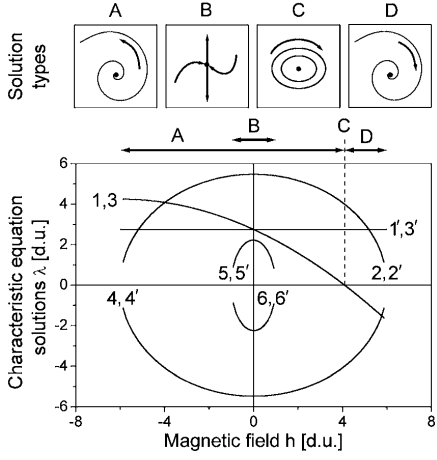}
\caption{ Magnetic field dependence of the characteristic equation
solutions: real (curves~1 and 3) and imaginary (curves~2 and 4)
parts of $\lambda_1$, real parts of $\lambda_2$ (curves~5 and 6).
Stationary states: curves 1 -- 6, equilibrium states: curves
$1^\prime$ -- $6^\prime$. The curves 1,3 and $1^\prime$,
$3^\prime$ are multiplied by the factor of 100.}
\end{figure}

The solutions of the characteristic equations for the equilibrium [(\ref{Eq36}),
(\ref{Eq37})] and stationary [(\ref{Eq29}), (\ref{Eq33}), and (\ref{Eq34})]
states are plotted in Fig.~4 as a function of  $h$. The solutions corresponding
to the stationary state ($\theta_{01}$, $\varphi_{01}$) are complex conjugated
(curves 1, 2 and 3), whereas the ones corresponding to ($\theta_{02}$,
$\varphi_{02}$) are real numbers of opposite signs (curves 5 and 6). It is
worth noting that the real part $\lambda_r$ (curves 1 and 3) for the stationary
state ($\theta_{01}$, $\varphi_{01}$) can be positive (area A), negative (area
D) or equal to zero (point C).

\begin{figure}\label{Fig5}
\includegraphics[scale=1.2]{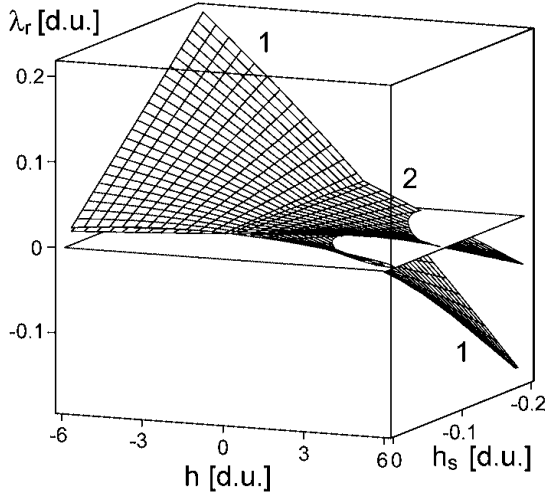}
\caption{ The real part of the characteristic equation solution as
a function of $h$ and $h_s$: 1~--~stationary state,
2~--~equilibrium state.}
\end{figure}

As concerns solutions of the characteristic equation for the
equilibrium state (\ref{Eq36}), we note that the curves $5^\prime
$ and $6^\prime $ in Fig.~4 practically coincide with the
solutions for the stationary states ($\theta_{02}$,
$\varphi_{02}$) (curves 5 and 6). For the state (\ref{Eq37}), only
the imaginary parts of the solutions (curves $2^\prime $ and
$4^\prime $) coincide with the imaginary parts describing the
state ($\theta_{01}$, $\varphi_{01}$) (curves 2 and 4), whereas the
real parts of these solutions differ significantly (curves
$1^\prime $, $3^\prime $ and 1, 3 in Fig.~4 are scaled by the
factor of 100 for better visual presentation). It is important
that for the given values of the control parameters, the real part
of (\ref{Eq37}) does not depend on $h$. However, if the control
parameters are changed in a sufficiently wide range, the real
parts of the solutions (\ref{Eq37}) and ($\theta_{01}$,
$\varphi_{01}$) behave similarly (Fig. 5). The difference between
them is seen in the intersection of the surfaces
$\lambda_{ri}=f(h,h_s),\; (i=e,\, 0)$ with the $\lambda_r=0$
plane, which for the stationary state takes place for weaker
magnetic fields.

According to the classification of specific points describing the motion in the
vicinity of equilibrium states,\cite{aRef39} the solution ($\theta_{02}$,
$\varphi_{02}$) corresponds to a saddle point (Fig.~4, B) with two attracting
equilibrium states. The second stationary state ($\theta_{01}$, $\varphi_{01}$)
has a broader variety of possible motion types: unstable focus (Fig.~4, A),
stable focus (Fig.~4, D), and stable center (Fig.~4, C). It is interesting to
note that the point, where the stable center appears is located on the boundary
between the intervals of stable and unstable focuses. Therefore, the condition
of reaching the stable center, $\lambda_r(h_b)=0$, determines the bifurcation point
$h_b$ (the boundary of neutral stability\cite{aRef31}) as
\begin{eqnarray}
\label{Eq53} h_{b0}=\frac{h_s}{\alpha}+\sqrt{\left(\frac{h_s}{\alpha}
\right)^{\! 2} + \left( 1+ \frac{h_p}{2}+ \sqrt{\varpi} \right)^{\! 2} -
\frac{\varpi}{4}},
\end{eqnarray}
where $\varpi \equiv h_p^2 - 4 h_s^2$.

As follows from Eq.~(\ref{Eq53}), with  $h_s$ decreasing from zero
to $h_s=-h_p/2$, the value of $h_b$ diminishes monotonously from
$h_{b0}^{max}=\sqrt{(1+h_p)(1+2h_p)}$ to
$h_{b0}^{min}=-(h_p/2\alpha)+ \sqrt{(h_p/2\alpha)^2+(1+0.5\,
h_p)^2}$ (see Fig.~6). At the same time, $h_{b0}$ shows a
non-linear increase with the anisotropy $h_p$. It is worth noting
that for the equilibrium state (\ref{Eq37}), the neutral stability
field boundary $h_{be}$ (Fig.~6, dashed lines) is determined from
the expression,
\begin{equation}
\label{Eq54}h_{be}=-\frac{(1+h_p)(1+2 h_p)}{h_s \alpha^{-1} + 1 +h_p},
\end{equation}
satisfying also the condition $h_{be} > h_{b0}$. In this way,
(\ref{Eq53}) and (\ref{Eq54}) determine the system stability
limits, as the corresponding states are stable when $h>h_{be}$ ($i
= 0,e$) and unstable in the opposite case.

\begin{figure}\label{Fig6}
\includegraphics[scale=1]{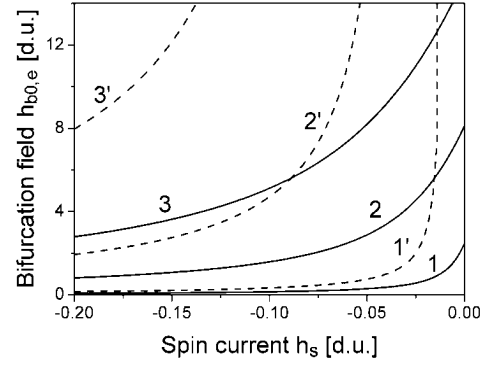}
\caption{ Bifurcation diagram of neutral stability for stationary
(solid curves) and equilibrium (dashed curves) states for $\alpha
= 0.005$ and different $h_p$: 1 and $1^\prime$ correspond to
$h_p=1$; 2 and $2^\prime$ correspond to $h_p=5$; 3 and $3^\prime$
correspond to $h_p=10$.}
\end{figure}

\begin{figure}\label{Fig7}
\includegraphics[scale=1.2]{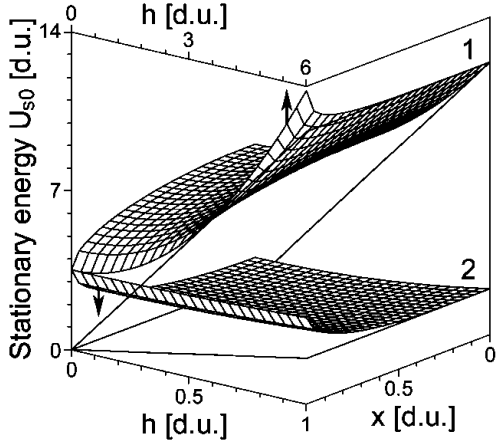}
\caption{ Dependence of the stationary energy on $h$ and $x \equiv | 2 h_s /
h_p |$: 1~-~$U_{s01}$, 2~-~$U_{s02}$.}
\end{figure}

The two-dimensional plot of the energy of both $U_{s01}$ and $U_{s02}$
stationary states, calculated using (\ref{Eq31}), is presented in Fig.~7 for $0
\le |h| \le 1$ (state ($\theta_{02}$, $\varphi_{02}$)) and for $0 \le |h| \le
6$ (state ($\theta_{01}$, $\varphi_{01}$)); $0 \le x \equiv \left|2h_s / h_p
\right| \le 1$. For $h=0$ both surfaces merge, taking the energy value
$U_{s0}(h=0)=1+0.5h_p$. With increasing $h$, the energy of the low-energy state
$U_{s02}$ increases slightly, but this increase is faster for larger values of
the spin current $h_s$. Contrary, the changes of the high-energy state
$U_{s01}$ with increase of either $h$ and $h_s$ are significant, non-linear,
and lead to much greater difference in the energy of the states $U_{s01}$ and
$U_{s02}$. Therefore, we will focus below on the non-stationary magnetization
dynamics, corresponding to the excited state $U_{s01}$.

\begin{figure}\label{Fig8}
\includegraphics[scale=1.2]{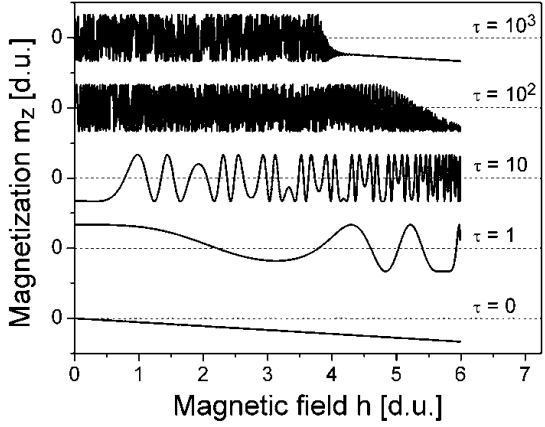}
\caption{ Dependence of $m_z$ on external magnetic field for the
stationary state at different observation times.}
\end{figure}

\begin{figure}\label{Fig9}
\includegraphics[scale=1.2]{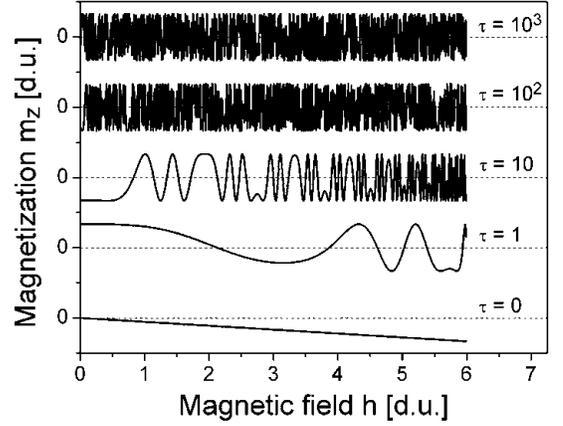}
\caption{ Dependence of $m_z$  on external magnetic field for the
equilibrium state at different observation times.}
\end{figure}

The time and field behavior of the longitudinal magnetization component $m_z$
under small perturbations are illustrated in Figs.~8 and 9 for the stationary
($\theta_{01}$, $\varphi_{01}$) and equilibrium ($U_{04}$) states, using
formulas (\ref{Eq33}), (\ref{Eq34}) and (\ref{Eq44}). At the initial moment
($\tau=0$) and after a short period of time ($\tau=1$), the changes of the $m_z$
component with the increasing magnetic field are equal for both stationary and
equilibrium states, but with a further increase of the observation time, the
character of these curves becomes different, at first slightly ($\tau = 10$)
and then significantly ($\tau \ge 10^2$). It is worth noting that the
oscillations become chaotic, causing $m_z$ to take random values in the range
of $-1 \le m_z \le 1$. It is important that for $\tau \ge 10^3$ and $h>4.2$
(Fig.~8), the perturbation of the state ($\theta_{01}$, $\varphi_{01}$) relaxes
to the linear $m_z = f(h)$ dependence, comparable with Fig.~3 (curve 1), while the
relaxation of equilibrium state (\ref{Eq37}) takes place for $\tau > 10^4$.

Therefore, under the influence of small perturbations, the high-energy
stationary and equilibrium states of the ferromagnetic system in magnetic field
are characterized with the complex temporal dynamics, which can be investigated in
detail using the numerical methods to solve Eqs.~(\ref{Eq5}).

\subsection{Results of numerical modelling}

The type of spin orientation in the system varies significantly
with a change of the applied magnetic field. Our investigation
reveals the presence of several oscillation regimes depending on
the value of the applied magnetic field $h$. The set of differential
equations (\ref{Eq5}) was solved numerically using the Runge-Kutta
method of the fourth order.\cite{aRef44} This allows to  observe
the motion of the magnetization vector in the three-dimensional
space, forming a phase portrait of the system. To determine the
effect of magnetic field on switching between the states
$m_z=\pm1$, the magnetization vector was assumed to point down at
the initial moment, i.e., $m_x=m_y=0$, $ m_z=-1$.

\begin{figure}\label{Fig10}
\hspace*{-0.5cm}
\includegraphics[scale=1.1]{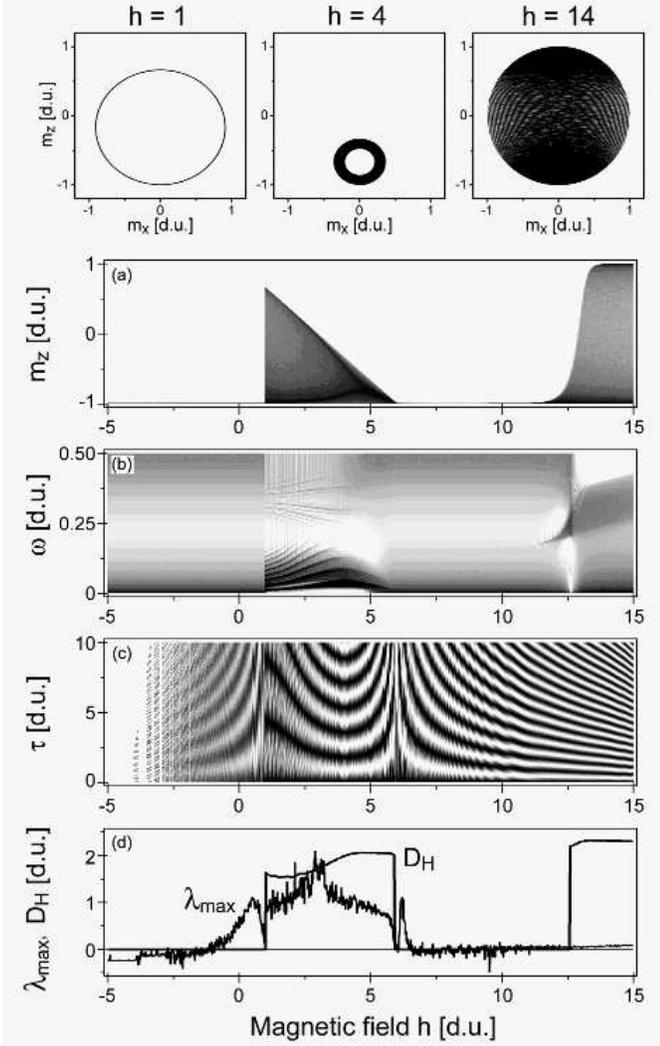}
\caption{ Phase portraits and main system characteristics {\it vs}
applied magnetic field: (a) $m_z$ density plot, (b) $m_z$ power
spectral density $S(\omega)$, (c) evolution of trajectory tracing
curve, and (d) Hausdorff dimension $D_H$ and maximum Lyapunov
exponent $\lambda_{max}$.}
\end{figure}

The most characteristic phase portraits of the system, shown in Fig.~10, were
obtained for $h_s=-0.03$, $-5 \le h \le 15$, $\alpha = 0.005$, and $h_p=5$. The
density plot of $m_z$ component versus applied field (Fig.~10(a)) features
darker areas corresponding to the most frequent $m_z$ values at a given
magnetic field. The field dependence of the power spectral density $S(\omega)$
of the component $m_z$ is given in Fig.~10(b), showing the peaks as dark areas.
The precession velocity can be estimated from the trajectory tracing curve
evolution,\cite{aRef45, aRef46} shown in Fig.~10(c). Here the dark areas
correspond to the moments of time $\tau$, when the vector connecting two
consecutive phase points becomes parallel to the one between two initial
points; the distance between two black stripes $\Delta\tau$ is inversely
proportional to the velocity of the phase point. The Hausdorff dimension $D_H$
of the phase portrait and maximum Lyapunov exponent $\lambda_{max}$ are shown
in Fig.~10(d) for different values of the applied field.

For magnetic fields $h<1$ the phase point remains in the vicinity
of the ground state $m_z=-1$ during all the observation time,
which is clearly seen at the density plot as the dark horizontal
bottom line. The trajectory tracing curves allow to discern a negligibly
small movement of the phase point characterized by a period
increasing with the applied magnetic field. The tendency to
instability is also notable from the $\lambda_{max}$ curve
reaching positive values. When the applied field $h$ surpasses
$h=1$, the system switches to the magnetization precession mode,
characterized by the periodic phase point orbit in $m_x, m_z$
plane (Fig.~10, $h=1$). The corresponding place of the density plot
shows two darker branches, marking the upper and lower limits of
$m_z$ values, which are contracting towards $m_z=-1$ with
increasing magnetic field. Darkening of almost homogeneous gray
area between these branches at higher $h$ is caused by the limit
cycle stability loss, visually observable as a phase trajectory
broadening (Fig.~10, $h$=4). At a certain merging point of $h
\approx 5$, the darker branches join together and the phase
portrait of the system turns to the focus, i.e., the magnetization
precession converges to a value close to the state $m_z=-1$.

It is worth noting that the $S(\omega)$ plot reveals several peaks
corresponding to the harmonic oscillations in the limit cycle
mode. The number of these peaks diminishes upon reaching the
magnetic fields above which the phase portrait of the system turns
into a focus. As can be deduced from the trajectory tracing plot,
the initial shrinking of the limit cycle oscillation mode is
accompanied by the precession frequency increase almost up to the
merging point. The movement of the phase point becomes then slower
for higher magnetic fields. The limit cycle stability loss and the
corresponding increase of the phase portrait density could be
clearly seen from the Hausdorff dimension curve, featuring a rapid
increase from 1.55 to 2.05 before the merging point and almost no
changes above it. The Lyapunov exponent has a noisy maximum at
$h\approx 3$, revealing a return to the stable state upon
switching to the converging magnetization precession mode.

It is worth noting that in the whole range of applied field,
$6<h<11$, the magnetization vector is restricted to the closest
vicinity of its ground state $m_z=-1$. At the same time, as it can
be deduced from the trajectory tracing plot, the frequency of the
phase point precession increases, and starting from $h\approx
11.1$ the system begins to tend to the upper state with $m_z=1$
through the magnetization precession. The phase portrait
corresponding to this oscillation mode represents a sphere formed
by a tight spiral with Hausdorff density $D_H=2.3$ (Fig.~10,
$h$=14). It is worth noting a rather stable nature of such
precession, described with almost zero values of $\lambda_{max}$.

A similar magnetization dynamics is also observed for different
values of $h_s$, as it can be seen from Fig.~11, presenting
several $m_z$ density plots. In all the cases, the system under
consideration does not leave the vicinity of the ground state
$m_z=-1$ until the magnetic field overcomes the value $h=1$. Above
this point, it is possible to observe a magnetization vector
precession to the state $m_z=1$ for $h_s=0$ and $h_s=-0.01$, while
for higher negative spin current values, the system is switched to
the limit cycle mode. From now on, the system evolution follows
the same scenario with the limit cycle contracting into a focus.
Depending on the spin current value, the system remains in the
vicinity of the bottom ground state for different ranges of the
applied field $h$. Starting from some threshold field, which
increases with the absolute value of $h_s$, one can observe the
switching of the system from the bottom to the upper ground state.

\begin{figure}\label{Fig11}
\includegraphics[scale=1]{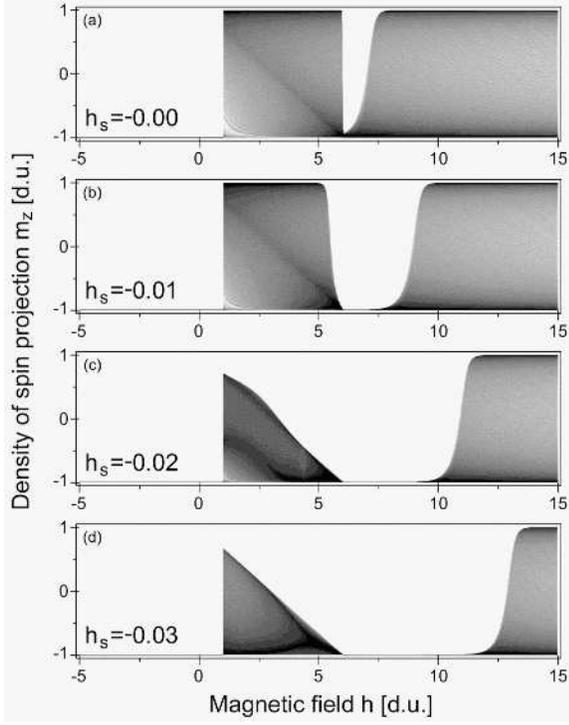}
\caption{ The system evolution with the applied magnetic field
$h$: $m_z$ density plots for different values of the spin
current.}
\end{figure}

\begin{figure}\label{Fig12}
\includegraphics[scale=1]{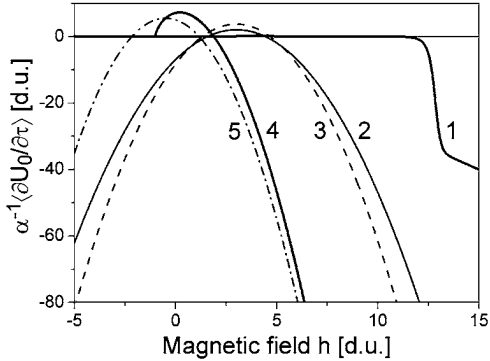}
\caption{Dependence of $\alpha^{-1}\langle \partial U_0 / \partial
\tau \rangle$ on the applied magnetic field for different
averaging procedures: 1 - according to (\ref{InsEq53}), 2 - from
the formula (\ref{InsEq49}), 3 - using (\ref{InsEq50}), 4 - from
(\ref{InsEq53}), 5 - according to the formula (\ref{InsEq51}).}
\end{figure}

\begin{figure}\label{Fig13}
\includegraphics[scale=1]{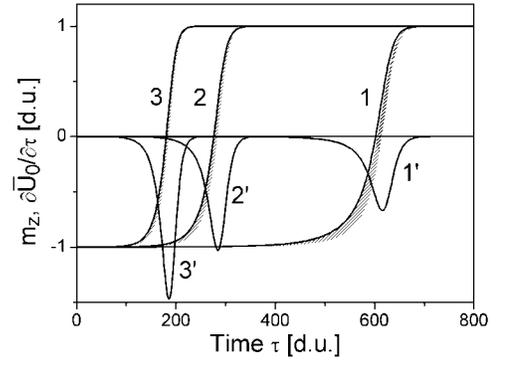}
\caption{Time evolution of $m_z$ (curves 1, 2, 3) and $\partial
\overline{U}_0 / \partial \tau$ (curves 1', 2', 3'), calculated
for $h_s=-0.03$ and different applied magnetic fields: 1,1' for
$h=11$; 2,2' for $h=13$; 3,3' for $h=15$.}
\end{figure}

\begin{figure}\label{Fig14}
\includegraphics[scale=1]{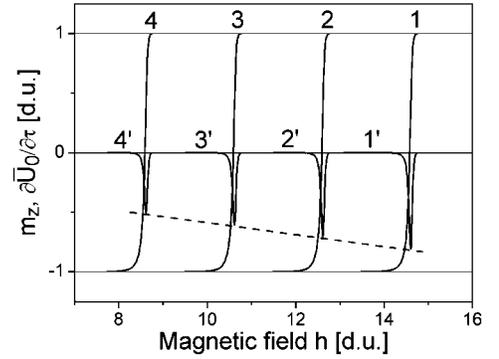}
\caption{Magnetization component $m_z$ (curves 1-4) and $\partial
\overline{U}_0 / \partial \tau$ (curves 1'-4') as a function of
magnetic field for $\tau = 800$ and different spin currents: 1,1'
- for $h_s=-0.05$; 2,2' - for $h_s=-0.04$; 3,3' - for $h_s=-0.03$;
4,4' - for $h_s=-0.02$.}
\end{figure}

\begin{figure}\label{Fig15}
\includegraphics[scale=1]{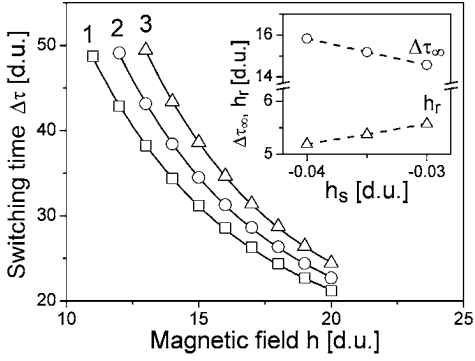}
\caption{Dependence of switching time $\Delta\tau$ on $h$ for
different spin currents: 1 - for $h_s=-0.030$, 2 - for
$h_s=-0.035$, 3 - for $h_s=-0.040$. Behavior of the parameters
$\Delta\tau_\infty$ and $h_r$ with spin current $h_s$ is given in
the inset.}
\end{figure}

In the framework of our description, we have performed the
averaging of the expression (\ref{Eq45}) over time according to
Ref.~[\onlinecite{aInsRef35}]:
\begin{equation}
\label{InsEq53} \left\langle \frac{\partial U_0}{\partial \tau} \right\rangle =
\lim \limits_{T \to \infty} \frac{1}{T} \int \limits_0^T \frac{\partial
U_0(\tau)}{\partial \tau} d\tau,
\end{equation}
where the value $\partial U_0 / \partial \tau $ was calculated for
$\theta (\tau)$ and $\varphi (\tau)$, determined from the
numerical solution of  Eqs.~(\ref{Eq5}) using the Runge-Kutta
method, and changing the condition $T \to \infty$ in
(\ref{InsEq53}) to $T \to T_{max}$, where $T_{max}$ was selected
from $\langle \partial U_0 / \partial \tau \rangle$ saturation
condition.

>From the comparison of the curve 1 in Fig.~12, calculated
according to formula (\ref{InsEq53}), with the curves 2 and 3
calculated from (\ref{InsEq49}) and (\ref{InsEq50}), follows that
there are significant quantitative and qualitative differences
between the time (\ref{InsEq53}) and phase (\ref{InsEq47}),
(\ref{InsEq48}) averaging procedures. However, for the Sun's model
$(\theta \ll 1)$ with  $h>0$, the results of phase averaging
(Fig.~12, curve 5) calculated according to the formula
(\ref{InsEq51}) approach asymptotically the results of time
averaging (curve 4, obtained for $\theta_0 = 1$) with increasing
external magnetic field. Such a difference between the phase and
time averaging of Eq.~(\ref{Eq12}) for the case of arbitrary
$\theta$ and for the particular case of $\theta \ll 1$ are caused
by the fact that the averaging procedures (\ref{InsEq47}) and
(\ref{InsEq48}) consider incorrectly the weighting factors for the
polar angle $\theta$.  One of the other evidence supporting this
point is the expression for charge current $I_c$, determined on
the base of (\ref{InsEq49}) (or (\ref{InsEq50})) from the
condition $\langle
\partial U_0 / \partial \tau \rangle = 0$,\cite{aRef20}
\begin{eqnarray}
\label{InsEq54} I_c=\frac{2e}{\hbar \eta}\; a^2 l_m M_s H_k \\ \times
\frac{\alpha}{h}\left[ h^2 + \frac{1}{4}\left(1+\frac{h_p}{2}
\right)^2+\frac{5h_p^2}{32}\right], \nonumber
\end{eqnarray}
which reveals an unexplainable divergence at $h \to 0$. At the
same time, expression (\ref{InsEq52}) practically corresponds
to the averaging only over the angle $\varphi$, which -- taking
into account a good correlation of curves 4 and 5 (Fig.~12) --
seems to be the most appropriate one. This result allows to
simplify significantly initial equations (\ref{Eq4}), (\ref{Eq5}),
and (\ref{Eq12}) by averaging their right-hand sides over
$\varphi$. As a result, instead of (\ref{Eq5}) one obtains
\begin{eqnarray} \label{InsEq55} \frac{\partial m_z}{\partial
\tau}= \alpha (1+0.5h_p)(1-m _z^2) (m_z+\beta_p), \\
\frac{\partial \overline{\varphi}}{\partial \tau} = -(1+0.5h_p)m_z
+ \alpha h_s - h. \nonumber
\end{eqnarray}

Instead of (\ref{Eq4}) and (\ref{Eq12}) it is possible to get
correspondingly
\begin{equation}
\label{InsEq56} \overline{U}_0 = (1+0.5h_p)(1-m_z^2)-2hm_z,
\end{equation}
and
\begin{equation}
\label{InsEq57} \frac{\partial \overline{U}_0}{\partial \tau} = -2 \alpha
(1+0.5h_p)^2(1-m_z^2)(m_z+\beta_p)(m_z+\gamma),
\end{equation}
with $m_z=\cos \theta$, $\beta_p \equiv (h+h_s/ \alpha) / (1+0.5h_p)$ and
$\gamma \equiv h / (1+0.5 h_p)$.

The ordinary differential equations (\ref{InsEq55}) are of
separable-variable type and allow to find solutions for the
averaged azimuthal angle $\overline{\varphi}$ at known $m_z(\tau)$
in the form
\begin{equation}
\label{InsEq58} \varphi(\tau)=\overline{\varphi}(0)-(h-\alpha h_s)\tau -
\left( 1+\frac12 h_p\right) \int\limits_0^\tau m_z(\tau)d\tau,
\end{equation}
where $\overline{\varphi} (0)$ is the initial value of the angle
(i.e., its phase).

In the particular case of $\beta_p = \pm 1$, $m_z(\tau)$ can be
found from the following transcendent equation:
\begin{eqnarray}
\label{InsEq59} \left[ \arctanh m_z \mp \frac{1}{1 \pm m_z}\right]\\ - \left[
\arctanh m_{z0} \mp \frac{1}{1 \pm m_{z0}}\right] = \mp 2 \alpha \,
(1+0.5h_p)\, \tau, \nonumber
\end{eqnarray}
where $m_{z0}$ is the initial value of the longitudinal
magnetization vector component. As follows from
Eq.~(\ref{InsEq59}), the sign change of the parameter $\beta_p$ is
equivalent to the simultaneous sign change of $m_z$ and $\tau$. It
is worth noting, that despite the case $\beta_p = \pm 1$ was
analyzed in detail\cite{aRef20} (see, for
example, Eq.~(19)) using numerical methods, the analytic equation
for $m_z(\tau)$ was not presented there.

For $\beta_p \ne \pm 1$,  $m_z(\tau)$ satisfies the following
transcendent equation:
\begin{eqnarray}
\label{InsEq60} \left( \frac{1+m_z}{1-m_z} \right)^{\beta_p}
(1-m_z^2)(m_z+\beta_p)^{-2} \nonumber
\\ = \left( \frac{1+m_{z0}}{1-m_{z0}}\right)^{\beta_p}
(1-m_{z0}^2)(m_{z0}+\beta_p)^{-2}\\ \times \exp \left[ -2 \alpha
(1+0.5h_p)(1-\beta_p^2)\tau \right]. \nonumber
\end{eqnarray}
It is easy to notice that expressions (\ref{InsEq58}) and
(\ref{InsEq60}) are generalizations of formula (\ref{Eq26})
for the case of $h_p \ne 0$.

As the case $\beta_p = \pm 1$ (Eq.~(\ref{InsEq59})) was already
investigated in detail,\cite{aRef20} where it was used for determination of the
threshold field $\left| h_{ac}^{(\pm)}\right|$ for the switching from parallel
to antiparallel $m_z$ state, we will consider below only the switching
peculiarities defined by Eq.~(\ref{InsEq60}).

In Figs. 13 and 14 we present the time and field dependence of
$m_z$ and $\partial \overline{U}_0 / \partial \tau$, calculated
according to Eqs.~(\ref{InsEq60}) and (\ref{InsEq57}),
respectively. The dashed areas in Fig.~13 correspond to
$m_z(\tau)$ values obtained by the numerical integration of the
system using the Runge-Kutta method. As one can see, for different
values of $h$, $m_z(\tau)$ determined from Eq.~(\ref{InsEq60})
forms the outline of numerically calculated magnetization vector
projection. It is worth noting that the good correlation between
the analytical and numerical results can be also observed for the
function $m_z(h)$ for different values of spin current $h_s$ (Fig.
14). The latter proves that the proposed simplification of initial
equations (\ref{Eq4}), (\ref{Eq5}), and (\ref{Eq12}) by averaging
over $\varphi$ is correct. The dashed curves in Figs.~13 and 14
correspond to time- and magnetic field-dependent values of
$\partial U_0 / \partial \tau$, calculated on the base of
Eqs.~(\ref{InsEq57}) and (\ref{InsEq60}). Our analysis has shown
that these curves can be approximated well with the Gaussian
distribution.\cite{aInsRef34} The values $m_{zm}(h, h_s, h_p,
\alpha)$ corresponding to the minimum of $\partial U_0 / \partial
\tau$, can be found by solving the cubic equation
\begin{equation}
\label{InsEq61} m_{zm}^3+\frac{3}{4}(\beta_p+\gamma)m_{zm}^2+
\frac{1}{2}(\gamma\beta_p-1)m_{zm} -\frac{1}{4}(\beta_p+\gamma)=0,
\end{equation}
which can have one or three real solutions, of which only those
satisfying the condition $-1\le m_{zm} \le 1$ should be taken into
account.

As follows from Fig.~13, at the fixed values of $h_s$, $h_p$ and $\alpha$, the
minimum value of $\partial \overline{U}_0/
\partial \tau = f(h)$ decreases exponentially with time, while the
similar minimum of $\partial \overline{U}_0/
\partial \tau = \psi(h_s)$ at fixed $\tau$, $h_p$, and $\alpha$ grows linearly
with the increase of $h_s$ (Fig. 14). It is important to emphasize
that in the latter case the value of applied magnetic field, at
which $m_z$ turns to zero, decreases linearly with increasing
$h_s$.

As one can see from Figs. 13 and 14, the switching time between
the ground states $m_z=-1$ and $m_z=+1$ has the same order of
magnitude as the half-width of the $\partial \overline{U}_0 /
\partial \tau$ peak. This half-width remains practically unchanged
with $h$ at the fixed $\tau$, $h_p$, and $\alpha$  (Fig. 14), but
depends significantly on the magnetic field with fixed $h_s$,
$h_p$, and $\alpha$ (Fig. 13). As it turns out (Fig. 15), the
half-width $\Delta\tau(h)$ of the peak $\partial U_0 / \partial
\tau = f(h,\tau)$ diminishes exponentially with increasing $h$
according to formula
\begin{equation}
\label{InsEq62} \Delta\tau=\Delta\tau_{\infty}+ (\Delta \tau_0 - \Delta
\tau_\infty)\exp (-h/h_r),
\end{equation}
where the parameters $\Delta\tau_{\infty}$, $\Delta\tau_0$ and
$h_r$ depend on $h_s$, $h_p$, and $\alpha$.  It is worth noting
that for the fixed applied magnetic field $h$, the increase of
$h_s$ leads to decrease of the switching time $\Delta\tau$. The
dependence of the parameters $\Delta\tau_{\infty}$ and $h_r$ on
the spin current $h_s$ are presented in the inset to Fig.~15. As
one can see from this figure, $\Delta\tau_{\infty}$ and $h_r$ are
characterized by a linear decrease and increase with growing
$h_s$, respectively. The results presented in Fig.~15 show that
the increase of the absolute value of spin current leads to
destabilization of the system, which now needs more time to switch
between the ground states $m_z=-1$ and $m_z=+1$. To the contrary,
the applied magnetic field has the stabilizing action, leading to
a decrease of $\Delta\tau$. It is also worth to emphasize that for
$h>25$, the influence of the applied magnetic field becomes
dominant compared to the role of the spin current $h_s$. It is
worth noting that formulas (\ref{InsEq60}) and (\ref{InsEq61})
allow to obtain analytical expressions for the parameters
(\ref{InsEq62}), making it possible to investigate in detail the
dependence of the switching time $\Delta\tau$ on $h, h_s, h_p$,
and $\alpha$.

\section{Conclusion}

We have studied the magnetization dynamics of a ferromagnetic
system subject to the spin-polarized current. We have used the
methods of non-equilibrium thermodynamics, which have been
developed to describe the self-organization processes.
\cite{aRef31}$^-$\cite{aRef34, aRef45, aRef46} Our results show
that the ferromagnetic system displays a complex dynamics of
instabilities with applied magnetic field and spin current. We
have demonstrated that the method can be used to describe the
dynamical properties of ferromagnetic nanostructures important for
spintronics applications.

The behavior of equilibrium and stationary states has been
investigated for a wide range of external parameters, without any
limits on the angular variables describing the system
magnetization. It has been shown that under certain conditions,
the system can be switched to the oscillation mode regime with
negligibly small damping. The phase portrait evolution has been
investigated in detail for different values of external magnetic
fields. The obtained results demonstrate a possibility to control
the operating modes of the ferromagnetic components in spintronic
devices by a proper choice of the ferromagnetic material with
appropriate easy-plane anisotropy and the damping coefficient. It
has been shown that the averaging of the initial system over the
azimuthal angle allows to obtain analytical expressions, which
make possible the detailed investigation of the switching
peculiarities of the longitudinal magnetization between the states
$m_z=-1$ and $m_z=+1$. The switching time has been analyzed in
dependence on the external magnetic field and spin current.

The obtained results proves the successful application of self-organization
methodology to solve theoretical problems of spintronics, which can yield new
quantitative and qualitative results.

\begin{acknowledgments}
This work is partly supported by FCT Grant POCI/FIS/58746/2004
(Portugal) and the Polish State Committee for Scientific Research
under Grants Nos. PBZ/KBN/044/P03/2001, 4~T11F~014~24 and
2~P03B~053~25. One of the authors (V.D.) thanks the Calouste
Gulbenkian Foundation in Portugal for support.
\end{acknowledgments}


\end{document}